\documentclass[a4, 11pt]{article}
\usepackage[textwidth=16cm]{geometry}
\usepackage{amssymb, amsmath}
\usepackage{graphicx}
\usepackage{color,cite,authblk}

\begin{document}

\title{Storage of electrical energy\\ {\small Batteries and Supercapacitors}}
\author{Trilochan Bagarti$^1$ and Arun M. Jayannavar$^2$\thanks{Email: jayan@iopb.res.in}}
\affil{$^1$Graphene Center, Tata Steel Limited, Jamshedpur-831001\\ $^2$Institute of Physics, Sachivalaya Marg, Bhubaneswar-751005}

\maketitle
\begin{abstract}
In this article we shall discuss the development of electrical storage system. Since the early days of 
electricity people have tried various methods to store electricity. One of the earliest device was a 
simple electrostatic capacitor that could store less than a micro Joule of energy. The battery has been the most 
popular in storing electricity as it has a higher energy density. We will discuss the development of 
batteries and their working principle. Although capacitors has never been thought of as a practical 
device for electricity storage, in the recent years there has been tremendous progress in building capacitors with
huge capacitance and it may soon be possible to use it as storage device. We shall discuss the technological
breakthroug in supercapacitor as a storage device.
\end{abstract}



\section*{Introduction}
\begin{figure}[!h]
  \begin{center}
    \includegraphics[width=0.2\textwidth]{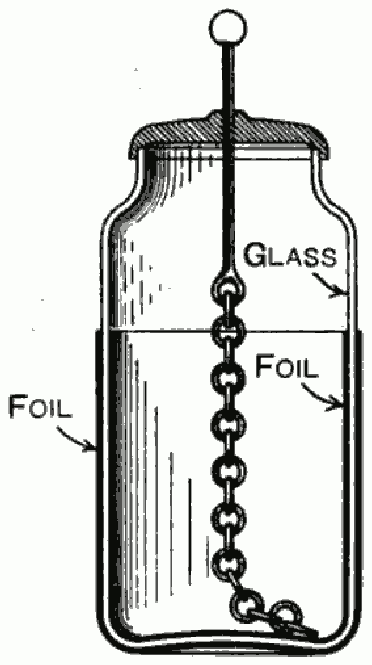}
    \includegraphics[width=0.2\textwidth]{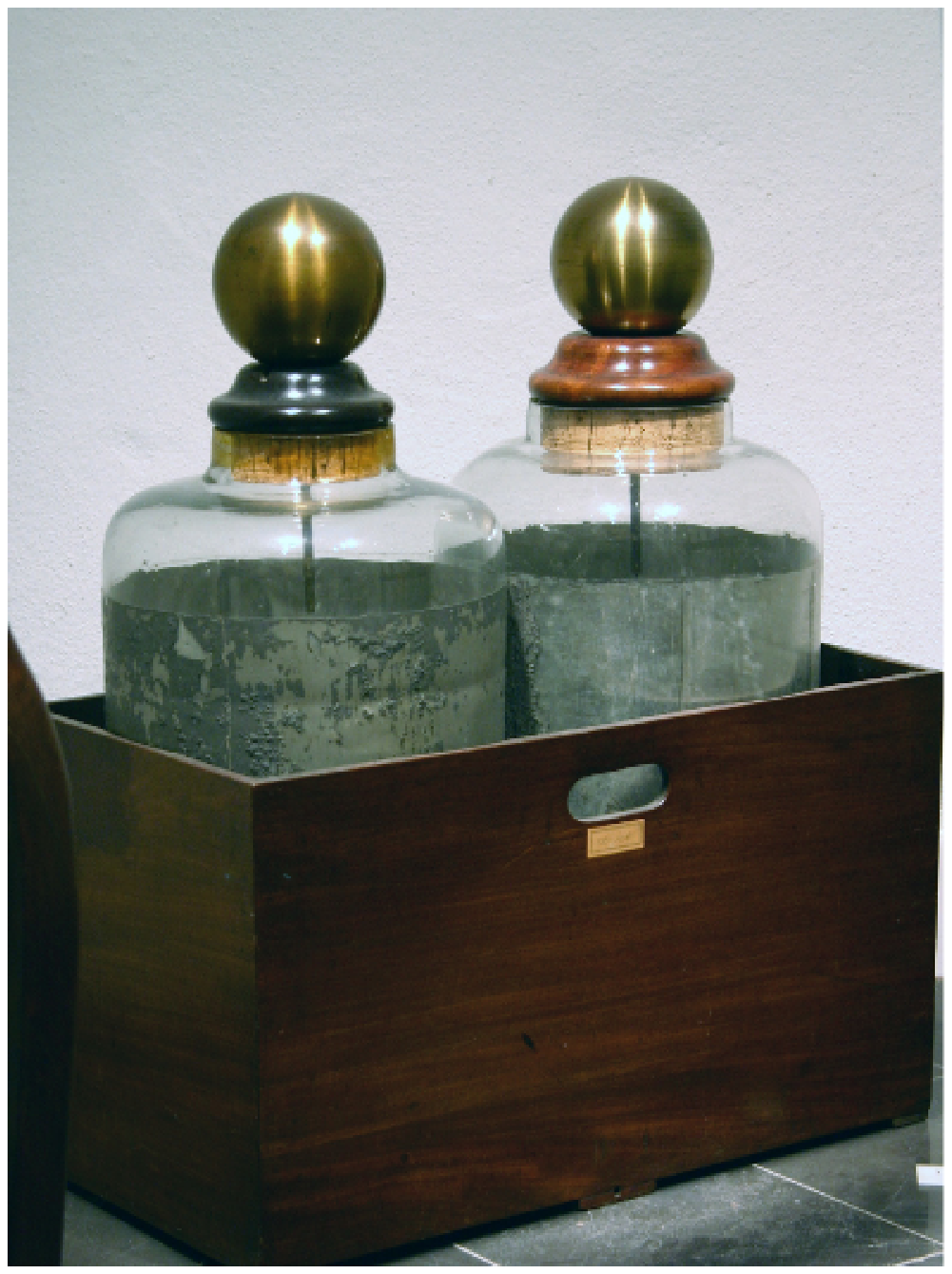}
  \end{center}
  \caption{The Leyden jar: Invented by Ewald Georg von Kleist and Pieter van Musschenbroek in 1745-46. (source: wikipedia.org)}\label{leyden}
\end{figure}

Electricity was know to many cultures as a mysterous force of nature since ancient times. The first scientific study 
was published only in 1600 by William Gilbert in "De Magnete" and until the begining of the ninteenth century any useful
application of electricity has not emerged. Today, we are all surrounded by huge number of devices 
that uses electrical energy to work. In these devices energy is stored in batteries in the chemical bonds of
the materials that is cleverly chosen to make a battery. However, more than a hundred years ago, when 
people were just begining to understand the nature of electricity with their intriguing experiments, 
mostly on static electricity, a very simple and useful device was invented - the Leyden jar. It was the first 
device that could store electrical energy. It consisted of a glass jar with conducting foil coated on the 
inner and outer surface and a metal electrode connecting the inner foil that projects through the
stopper at the mouth cf. Fig. (\ref{leyden}). The capacitor that is manufactured today has fundmantally 
the same designe except that it has become more efficient in storing energy. Today a capacitor can store
energy that is billions of times larger than that of a Leyden jar. How this has been achieved in the past 
several decades is most fascinating. In the following we shall discuss about batteries and capacitors and their uses
in electrical energy storage. We will see, how nanotechnology has made possible the construction of capacitors
with very large capacitance called supercapacitors. With recent breakthrough in nanotechnology it appears that the 
specific energy of supercapacitors may be comparable to certain batteries. 
\section{Electrochemical batteries}
\begin{figure}[!h]
  \begin{center}
    \includegraphics[width=0.15\textwidth]{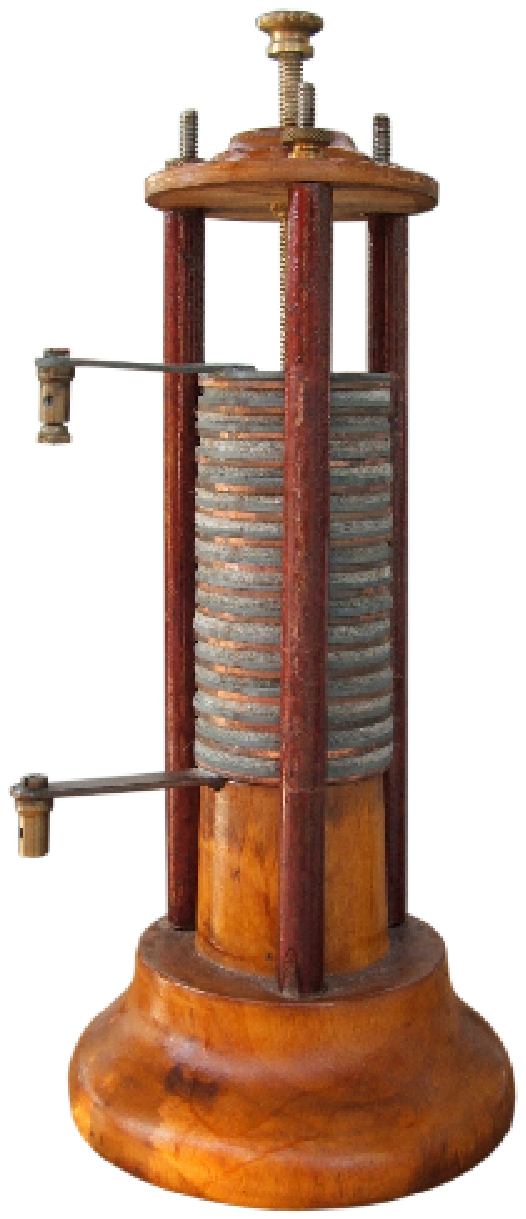}
    \includegraphics[width=0.35\textwidth]{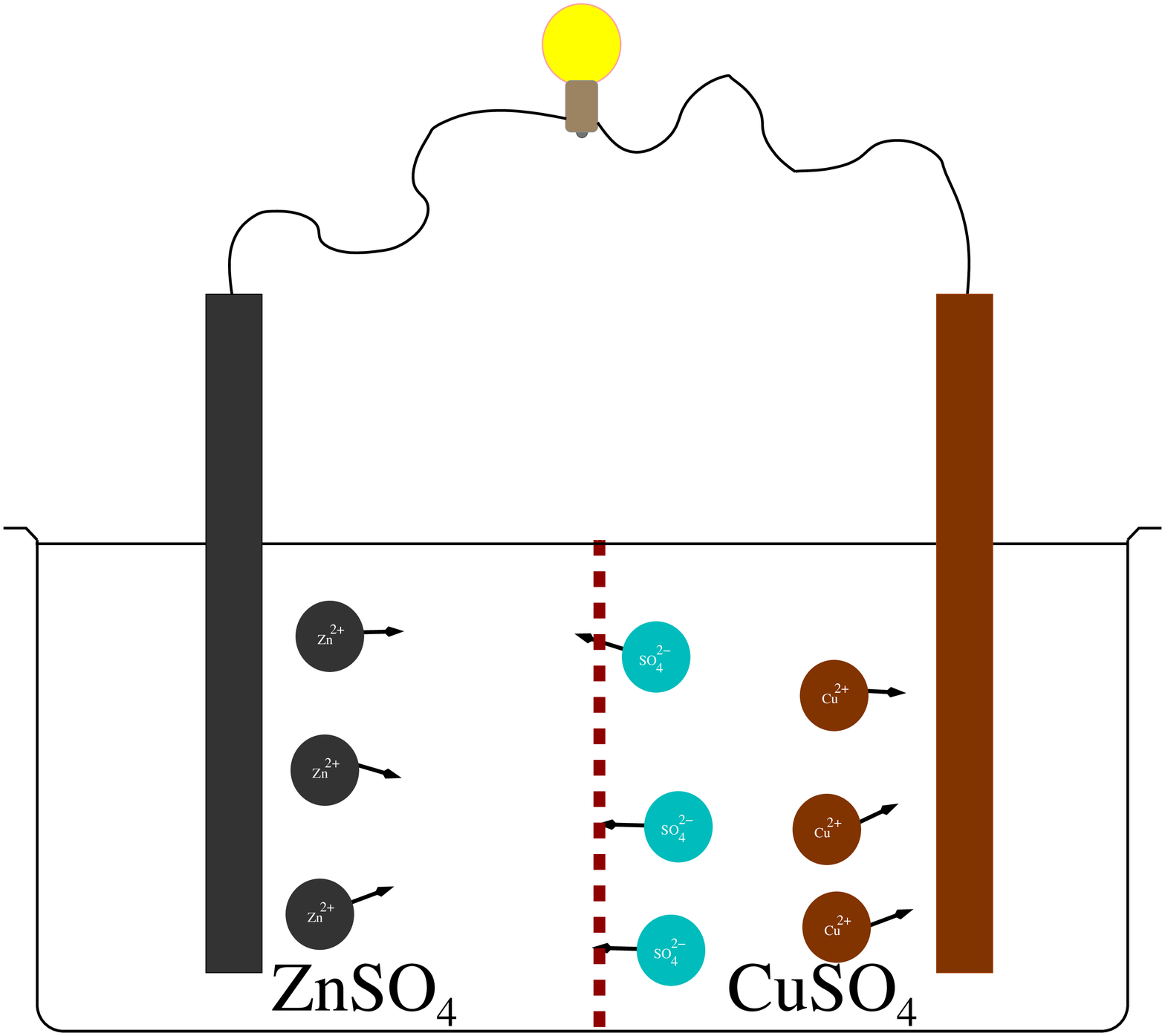}
  \end{center}
  \caption{What goes on in a voltaic pile? : Copper has a higher electron affinity as compared to that of Zinc. Electrons rushes
    through the wire from zinc to copper. An electric field is produced inside the media which drives the positive ions towards
    the copper and negarive ions towards zinc. This results in the formation of zinc sulfate in the left half cell and deposition
    of copper on the electrode in the right half cell. The semipermiable membrane allows the passage of sulfate ion from the right
    to the left half cell. (voltaic pile, source: wikipedia.org)}\label{voltaic-pile}
\end{figure}

Although, the term battery is used today for an electrochemical cell that converts chemical energy to electrical energy, it was actually 
used for a combination of Leyden jars by Benjamin Franklin. The first electrochemical battery consisted of zinc and copper disks 
separated by paper soaked in salt solution. It was invented by Alessandro Volta, a professor at University of Pavia, Italy in the year 
1800. A stack of several disks arrange alternately provided a large potential difference. Unlike a Leyden jar it was a source of continuous 
electrical current. It was also used to electrolyse water to obtain hyderogen and oxygen. The Voltaic pile is also sometime called the galvain 
cell after its inventor Luigi Galvani who used it to contract frog's leg which he named "animal electricity". Volta however believed that the
that electricity arises from the contact of metal electrodes! It was only after 34 years when Michel Faraday showed that electricity was generated
at the electrode surface due to the oxidation and reduction reaction. In Fig. \ref{voltaic-pile} (a) is shown a replication of the voltaic pile. The 
electrochemical processes in one of the cell is shown in the schematic Fig. \ref{voltaic-pile}(b).   
 
Here we still haven't yet discussed why the reaction proceeds in the way it does. 

\begin{figure}[!h]
  \begin{center}
    \includegraphics[width=0.3\textwidth, angle=-90]{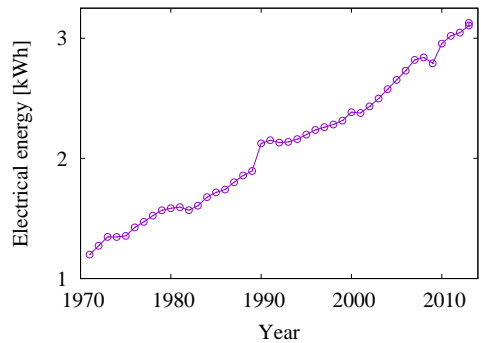}
  \end{center}
  \caption{World electricity consuption per capita (source: www.iea.org)}\label{electricity}
\end{figure}

\fbox{
\parbox[c]{10.0cm}{
~\\
{\bf Did you know?}\\
The energy consumption per capita in 2014 is 3.12 kW which is around three times that in the 1970 (cf. Fig. \ref{electricity}). 
It simply means that every year the consumption incesases by 44 Wh for each person! As oil and coal reserve continue to deplete 
and climate change becomes more severe we will be forced to use more renewable energy. In the next decade more people will be 
using electricity than fossil fuel. Technologies in future will give more freedom to use electrical energy as cost of electricty will reduce.

It reminds us of Michael Faraday, when in 1859, was asked by William Gladstone, British chancellor of the exchequer, questioned the 
usefulness of electricity, Faraday replied- "One day, sir, you may tax it."
~\\
} 
}

\section{Spontaneity and the Gibbs free energy}
It is not difficult to witnessing a spontaneous process in nature. Water flowing down a stream or smoke rising up from a chimney 
or the freezing of a lake during winter in cold contries are examples spontaneous processes. 
Have you ever wondered how a spoonful of sugar added to a cup of water spontaneously dissolves and vanishes into the water in a few minutes?
It disolves because by doing so the system goes to a lower free energy state. The second law of thermodynamics demands that the entropy
of an isolated system always increases. Here, in this particular example the Gibbs free energy of the syestem decreases. 
In all spontaneous processes that we observe, nature somehow tries to remain in the lowest free energy state.  
\begin{figure}[!h]
  \begin{center}
    \includegraphics[width=0.3\textwidth]{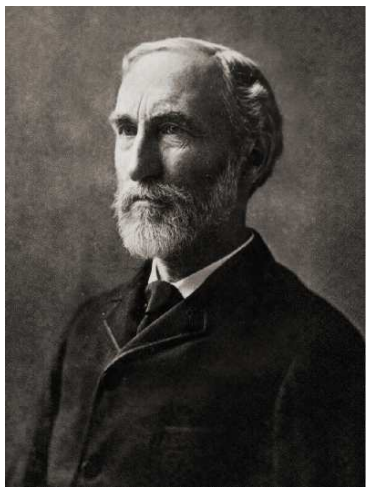}
  \end{center}
  \caption{Josiah Willard Gibbs (1839-1903), American theoretical physicist, one of the founders of statistical mechanics,
    introduced the concept of chemical potential.For a mixture of $N$ different species Gibbs gave a general
    expression for the change in energy $dE = T dS - pdV + \sum_i^N \mu_i dN_i$. The chemical potential $\mu_i = \partial E/\partial N_i$
    at constant entropy $S$, volume $V$ and particles $N_i$. Any open system undergoing a spontaneous process the Gibbs potential
    $G = H - TS$ attains a minimum. (source: wikipedia.org)}\label{jwgibbs}
\end{figure}

Consider a jar filled with white marbles upto say one third from the bottom and on top of it are filled up to the same height, 
 black marbels. So we have black marbles on top of white marbles - a very ordered state. 
Now shaking the jar well for some time will mix the marbels and we get a completely disordered state.  
In going from an ordered state to a disordered state we have just increased the entropy, i.e., the change in entropy $\Delta S >0$. 
At the molecular level, for the sugar molecules, the shaking is done by the thermal energy. The Gibb's free energy change of the system in
this example is  $\Delta G = - T \Delta S$ where $T$ is the temperature. In general the Gibbs free energy change
$\Delta G = \Delta H - T \Delta S$ where $\Delta H$ is the enthalpy or the heat absorbed at constant pressure. 

A spontaneous process in general is given by a positive entropy change as demanded by the second law. From processes at  constant temperature and 
at constant volume/pressure a spontaneous process may be defined by the Helmholtz and Gibbs free enargy change: 
\begin{align}
\Delta F < 0,  \text{ ~~~~(at constant volume, temperature)},  \\
\Delta G < 0, \text{~~~~(at constant pressure, temperature)}. 
\end{align}
At equilibrium the free energies attains a minima, hence $\Delta F =0$ and $\Delta G =0$ and there is positive free energy change for a non spontaneous
process.  

Before the invention of his voltaic pile, Prof. Volta performed hundreds of experiments with different pairs of metals and found that when
two dissimilar metal pieces are brought in contact with each other the two metal pieces aquire opposite charges.  Although at that time it was not
know that it was the electron that moves from one metal to the other creating a charge imbalance. Note that we can see great similarities between 
our marbels in the jar example and the electrons in the metals. The only difference here is that unlike marbles, electrons cannot be distinguish
with colors and they posses negative charge. 
\section{How batteries work?}
Consider the zinc and copper electrodes in a voltaic pile. The copper has a free electron density of $n_{Cu} = 8.59 \times 10^{28} / \text{m}^3$
and that in zinc is $n_{Zn} = 1.31 \times 10^{29} / \text{m}^3$. It easy to find these densities simply by considering the a unit cell of zinc
and copper and the number of electrons in it that can contribute to an electrical current. If the electrons had no charge the electrons would
simply move from zinc to copper and there would be an equilibrium density of $1.08 \times 10^{29}$ electrons in each of the metal pieces. Due
to the charge transfer there develops a potential difference across the contact surfaces called the contact potential or the Volta potential.
Let us assume that $n$ electrons move from the zinc to the copper.  Since the free energy change at equilibrium
$\Delta G = \Delta E + \sum_i \mu_i \Delta N_i$ vanishes. The first term is the change in the internal energy due to the contact 
potential and the second term is that due to the change in the number of particles of the $i^{th}$ species. Here, we just have one species
hence we should have, $0 = -en(\phi_{Cu} - \phi_{Zn}) + \mu_{Cu}n + \mu_{Zn}(-n)$. Note that $\Delta N$ is positive for copper and negative for Zinc.
Rearanging the terms gives us $\mu_{Cu} - e \phi_{Cu} = \mu_{Zn} - e \phi_{Zn}$. In general the following quantity may be constant across a contact surface
\begin{align}
\mu_0 + q \phi = \text{const},
\end{align}
where $\mu_0$ is the intrinsic chemical potential.

In the Voltaic pile there are two contact surfaces, i.e., the contact between the copper and the wet paper soaked in salt solution and similarly
between zinc and the wet paper. We know a salt solution does not conduct electricity by the free movement of electrons but instead it conducts by
oppositely moving ions. To understand how this happens we have a simplified Voltaic cell in Fig. \ref{voltaic-pile} (b). It consists of a the zinc
and copper electrode in zinc sulfate and copper sulfate solution which is separated by a semipermiable membrane in the middle which allows the
passage of sulfate ions only. As we saw earlier zinc loses a electrons through the wire and acquires a positive charge while copper electrod accepts
those electrons and becomes negatively charge. In the left and the right half cells we have the following reactions   
\begin{align}
Cu \rightarrow Cu^{2+}  + 2e^{-}, ~~~E^0 = 0.36 V, \nonumber \\
Zn \rightarrow Zn^{2+}  + 2e^{-},~~~ E^0 = -0.76 V. \nonumber
\end{align}
The negatively charged copper electrodes attracts $Cu^{2+}$ in the copper electrode and the $SO_4^{2-}$ sulfate ions get attracted towards the
positive charged zinc electrode. As a result zinc sulfate, $ZnSO_4$, is formed in the left half cell and $Cu$ is deposited on the right electrode.
Of course there has to be a movement on a sulfate ion pass through the membrane to balance the total charge to zero. In this process zinc is
oxidized by losing two electrons and copper is reduced by gaining the same number of electrons. This is called a redox reaction. The potential
difference can be obtained from the standard electrode potentials $E^0$of zinc and copper. i.e. from the half reaction described 
above the we have $E_{cell} = E^0_{red} - E^0_{ox} =0.34 - (-0.76) = 1.10 V$. The change in free energy $\Delta G = -n F E_{cell}$ where
$n=2$ is the number of moles of electrons transfered, $F = 98654$ Coulombs charge in a unit mole of electrons is the Faraday constant.
So the amount of energy stored in a battery which has anode and cathode weighing 65 g and the solution has 1 molar 
concentration of the salts will be $2.17 \times 10^{5}J$.

The Voltaic cell was a milestone in technological advancment. It started a new branch in chemistry-electrochemistry. Within months of the
invention of the invention of Voltaic pile several discoveries were made such as the splitting of water into hydrogen and oxygen, and the
isolation of varous alkali metals.
\section{The rechargable battery}
\begin{figure}[!h]
  \begin{center}
    \includegraphics[width=0.5\textwidth]{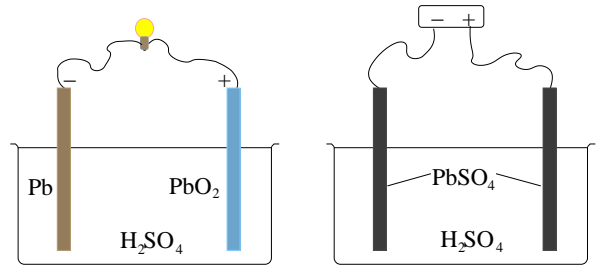}
  \end{center}
\caption{Working of a lead acid battery. Discharging cycle(left), and Charging cycle (right).}\label{lead-acid}
\end{figure}

The Voltaic cell is a one time use battery as the zinc electrode is consumed its potential drops to zero. This kind of cells are called primary cells.
In 1859, the French physicist Gaston Plant\'{e} invented the first rechargable battery, the lead-acid battery. After the battery is drained out applying 
a voltage with opposite polarity recharges the battery again. These kind of batteries are called secondary batteries.  To understand the working of a 
Lead acid battery we need to look at the redox reaction that takes place on the electrode surfaces during the charging and the recharging process.
During the discharging process we have 
\begin{align}
Anode: & Pb(s) + H_2SO_{4}(aq) \rightarrow PbSO_4(s) + 2H^{+}(aq) + 2e−, \nonumber \\
Cathode: & PbO_2 + 2H^{+} + H_2SO_4 + 2e^{-} \rightarrow PbSO_4(s) + 2H_2O(l). \nonumber
\end{align}
The working of a lead acid battery is show in Fig. \ref{lead-acid}. Note that during the dischrging process the lead-acid battery acts just a
voltaic cell where lead is oxidized at the anode and lead oxide is reduced at the cathode to form lead sulfate.
During the recharging process the opposite reaction takes place and the lead sulfate converts back to lead and lead oxide electrodes.
From the standard electrode potential we obtain a cell potential of $E_{cell} = 2.04$ Volts.
The lead acid brought a revolution is electrical storage device. Today there are a various other rechargable batteries in use. The lithium-ion
battery is the most advanced with the highest energy density. Although its devlopment goes back to the seventies, the first commercial battery
were introduced in 1991 by Sony and Asahi Kasei \cite{lib}. There are several types of lithium ion batteries such as lithium cobalt oxide (LiCO$_2$),
lithium manganese oxide (LiMn$_2$O$_4$, Li$_2$MnO$_3$), lithium nickel cobalt manganese oxide (LiNiMnCoO$_2$), lithium iron phosphate (LiFePO$_4$),
etc. The specific energy varries from 100-265 Wh/kg. In the lithium ion batteries the lithium ions move from the cathode to the anode and 
vice versa during the charging and discharging cycle in a reversible process. Due to this mechanism of the lithium ions moving back and forth
it is called "rocking chair" battery.  In general we can represent the over all process by 
\begin{align}
Li_x A + B \leftrightarrow Li_{x-y}A + Li_y B, 
\end{align} 
where Li$_x$ A is a compound that can readily accept and exchange lithium ions and B is another lithium accepting compound \cite{scrosati}.
Here the process going from left to right represent discharging and the reverse process represents charging cycle. 

There are various other rechargable batteries with varied specific energies\cite{shukla, shukla2}. Lithium ion has one of the highest
specific energies in electrical energy storage system. 

\section{What are supercapacitors?}
A typical supercapacitor may have a capacity of $100-1000$ Farad. Now, it is important to first understand how large is a Farad?
Let's construct a capacitor with this aluminium foil of area $10 \times 10 \text{cm}^2$ as electrodes separated by a piece of paper. 
If we connect the electrodes to a battery charge ($Q$) will be accumulated on the foils which is equal to the capacitance ($C$) times 
the applied potential difference ($V$). The capacitance of this capacitor is $C = \epsilon_r \epsilon_0 A/d$ where $A$ is the area of the foil, $d$ is 
thickness of the paper, $\epsilon_r$ is the dielectric constant and $\epsilon_0 = 8.85 \times 10^{-12} F/m^2$. Assuming that the paper
 has a dielectric constant $\epsilon_r = 2.5$ and thickness $d = 0.01 \text{mm}$ the capacitance can be computed to be 
$C = 2.2 \times 10^{-9} $ Farad or $C = 2.2$ nano Farad (nF). This is the capacitance of a typical Leyden jar! Charging it with a 
$V = 12$ Volt battery will therefore store energy $E = C V^2/2 = 0.16 \times 10^{-6}$ Joules of energy. Note that the energy 
stored in a capacitor is proportional to the square of the potential difference. Now to understand how big or small $0.16 \times 10^{-6} J$
 is lets consider a rubber band of length $l=10 cm$ and see how much energy is required to stretch it by 10 cm. The spring constant can be obtained
 by using the Hooke's law which says that stress ($\sigma$) is equal to the Youngs modulus ($Y$) times the strain ($\epsilon$) in a material, i.e.,
 $\sigma = Y \epsilon$. Lets assume that the rubber band has a diameter $d = 1$ mm and we have $Y = 0.0015$ giga Pascal
 \footnote{1 Pascal = 1 Newton/meter$^2$}, then the spring constant $k = Y A/l = 0.0015 \times 10^6 \pi d^2/ l  = 0.015 \pi$ Newton/meter.
 The energy stored in a spring which is stretched by an ammount $x$ in length is $E = k x^2/2$. So if we stretch the rubberband by $x = 1$ cm we
 require energy $E = 0.015 \pi (10^{-2})^2/2 = 0.23 \times 10^{-5}$ Joules. This is around ten times more energy that was stored in the Leyden jar.
 In other words we will need ten Leyden jar to stretch a rubber band by one centemeter!  

\begin{figure}[!h]
  \begin{center}
    \includegraphics[width=0.5\textwidth]{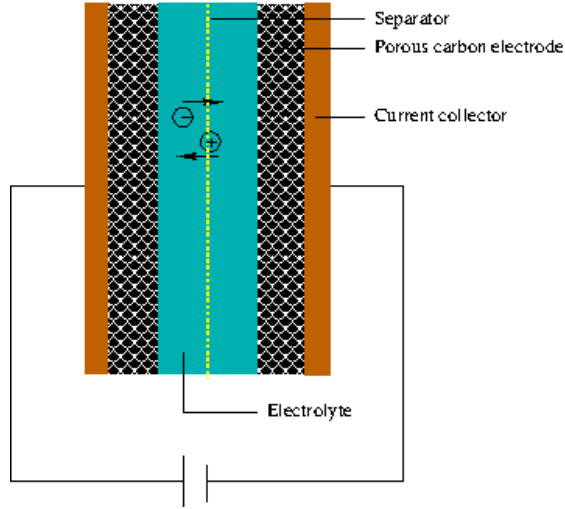}
  \end{center}
\caption{Constriction of a spercapacitor.}\label{edlc}
\end{figure}

 Our comparision of the energy stored is a capacitor with that of a battery tells us that  to use capacitors for electrical storage we need huge
 capacitance. The capacitance can be increased in two ways, one by increasing the area of the electrodes and two by increasing the dielectric constant.
 In the recent years there has been so much excitement and hope in utilizing the former by the use of advanced material such as graphene. We shall
 come back to it later but first see how a supercapacitor works. In 1957, H. Becker registered the first patent of an electrolytic supercapacitor that
 was constructed by the use of porous carbon electrodes and an electrolyte. It had a capacitance of around 6 Farad which he had no idea what gave rise
 to such a huge capacitance \cite{conway}. It was of course due to the large srface area of the electrodes that were responsibly for the rise in
 capacitance. However, the working of a supercapacitor is slightly different from that of the electrostatic capacitor. In Fig. \ref{edlc} we have a
 simple design of a supercapacitor. I consists of two electrodes with large surface areas usually made of carbon based materials such as activated
 charcoal, graphene, carbon nanotubes or a combinations of the three. In between that electrodes there there is an electrolyte and a porous membrane
 which allows the passage of ions. The porous electrode is in contact with a current collector usually metallic that has high electrical conductivity.
 When the electrodes of the capacitor is connected to an external source positive and negative ions in the electrolyte move in the opposite directions.
 They are accumulated on an electric double layer on the porous electrodes. The double layer consists of electrons and holes on the prorous carbon
 and oppositely charged ions on the surface separated by a thin layer of the solvent molecules. Now as we discussed earlier that the capacitance
 $C$ is inversely proportional to the distance $d$ between the electrodes i.e. $C =\epsilon_r \epsilon_0 A/d$, here we have a think monolayer separating
 the charges we have $d$ of the order of nanometers. This gives rise to a huge capacitance! This kind of capacitors are called the electric double layer
 capacitor.   

 Unlike batteries which can be charged and discharged for 1000 - 1500 times the supercapacitors practically lasts for ever since it has a
 charge-discharge cycle of $10^6$. The power density of a capacitor is also enormous due to very low equivalent series resistance which remains
 contant during discharge. Such large cycle is not possible in batteries since the electrodes degrade due to the redox reactions with every charge
 discharge cycle. Although the electric double layer capacitor stores energy by a completely non-Faradaic process there are however other kinds of
 capacitor called pseudocapacitors which utilizes a Faradaic charge transfere by the use of transition metal oxides such as RuO$_2$, IrO$_2$, FeO$4$, 
and MnO$_2$, titanium sulfides TiS$_2$. The specific capacitance of pseudocapacitors if larger than that of the electric double
layer capacitors but it has a lower specific power density. A list of commercially available supercapacitors may be found in \cite{scap}.
\section{Graphene in supercapacitors}
In the last couple of decades there has been tremendous progress in our understanding of the microscopic world.
Graphene is a atomically thin two dimensional sheet of carbon. It was discovered by Andre Geim and Kostantin Novoselov 
at the University of Manchester, UK in 2004 \cite{novo}. It consists of $sp^2$ bonded carbon atoms in a hexagonal lattice structure. 
It has exquisit electrical, thermal and mechanical properties. The surface area of one gram of graphene is 2630 $m^2$ and 
has excellent electrical conductivity. Theoretically the capacitance of a graphene based supercapacitor is 550 F/g. Almost all the 
supercapacitor available in the market uses some form of carbon based electrode with large surface area.
There is tremendous research in progress to achieve higher specific energy by using graphene based supercapacitors \cite{YuBin}.  

\section{Present status of electrical storage device}
\begin{figure}[!h]
  \begin{center}
    \includegraphics[width=0.6\textwidth]{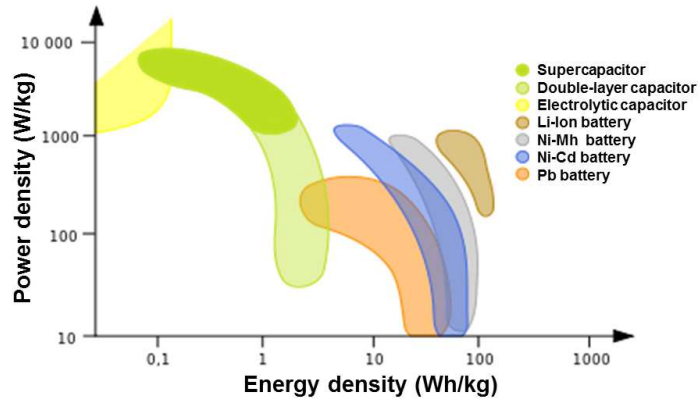}
  \end{center}
\caption{Ragone plot: comparision of energy storage devices. (source: wikipedia.org)}\label{ragone}
\end{figure}

In Fig. \ref{ragone}, we have a comparision of various kinds of electrical storage devices. The Lithium ion batteries has the highest specific 
energy of around of around 100-250 Wh/kg with a cycle life of around 1000 cycle. The Lead acid battery on the has a specific energy of 
30-40 Wh/kg \footnote{1 Wh = 3600 J} and remain in use and has one of the highest power density. Of all the batteries sold today the 
lead acid battery comprises around one half due to its cheaper raw materials and simpler technology. Recently the specific energy of supercapacitors 
appears to have reached an equivalent of the lead acid battery however it does not have long self life and hence cannot be used in place of the 
lead-acid batteries. However supercapacitor has been in use for in {\it capabus} in various countries that uses supercapacitor to store energy. 
At every stop the bus stops for a few minutes while the pasanger board/unboard it charges it supercapacitors through the charging point 
immediately. This energy is sufficient to run a few mile till it reaches the next bus stop. This technology has been used in Switzerland, Germany, 
UK, US and China. The technology is remarkable not oly because it is greener but also its uses almost 
half the energy required the capacitors have made the bus lighter. 

If we look at rechargable batteries that were used a decade ago we can recall that it took an awfull lot of time to recharge. Today the rechargable 
batteries are equiped with supercapacitors that enables it to charge is few minutes. These are call the hybrid batteries that uses a battery coupled to 
a supercapacitor. Interestingly, one of the earliest battery that used this technique is the hybrid lead acid battery, developed by CSIRO, Australia
in 2007. In China the CSR Zhuzhou Electric Locomotive has built a prototype of a light metro train that uses supercapacitors to run and uses no
other power supply \cite{csrz}.           
\section{Technologies to come in future}
The future of electrical storage devices looks not very coherent at the present time. Several promising
ideas is currently being pursued by researches around the world.
Even at present the Li ion battery is many fold expensive as compared other alternatives. In the immediate
future what we may see is a more powerful lithium battery. 
In this direction the technolology that could soon become a reality is the lithium-sulfur battery which
has a specific energy density as high as 500 Wh/kg. Recently, researchers
at the University of Alberta, Canada, have developed a silicon and graphene based lithium cell that has
a tenfold higher capacity than the currently available lithium batteries \cite{LiSi}. 
It would be essential to have technilogy that is not only advanced as well as affordaable. The sodium ion
battery looks very promising as it has a large cycle life of 2000 cycles although
the specific energy is low. 

Supercapacitors on the other hand will make possible technologies in the areas of wearable electronics,
energy scavenging, energy storage for microscopic machines and fast charging hybrid battery systems.
In a recent paper \cite{nitin} structural composites using reinforced carbon nanotubes electric double
layer capacitor were proposed which has multiple functionalality. Similarly, supercapacitors that can 
charge itself with the sun light has been reported which uses a graphene silver 3D foam electrodes \cite{libu}.  

\begin{figure}[!h]
  \begin{center}
    \includegraphics[width=0.5\textwidth]{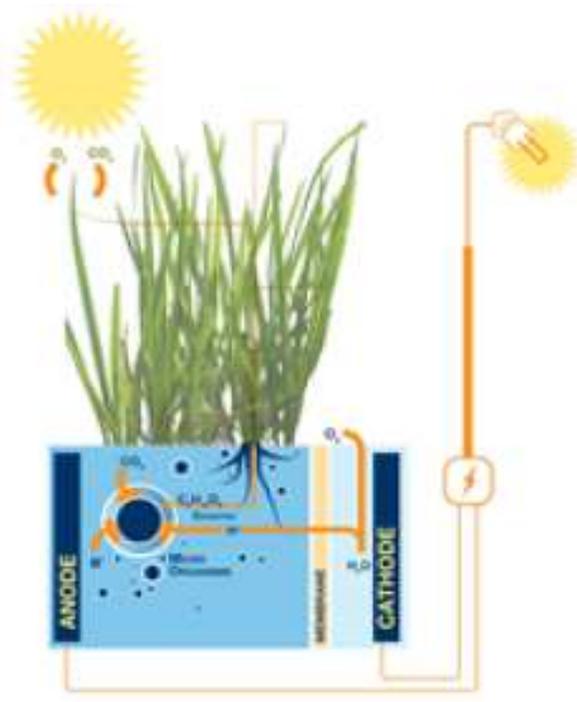}
  \end{center}
\caption{Plant mictobial fuel cell (source: Wageningen  University \& Research).}\label{pmfc}
\end{figure}
\fbox{
\noindent
\parbox[c]{10cm}{
~\\
{\bf Did you know?}\\
~\\
The bacteria present in the soil can be used to produce electricity. In 1911, M. C. Potter a professor in botany at the University of Durham,
used  S. cerevisiae a kind of yeast used in winemaking
to generate electricity and in 1931, B. Cohen constructed a microbial half cell that could generate 35 Volts with although a very small
current of 2 miliAmpere \cite{microbial}. Recently scientists at the  Wageningen University 
and the Wageningen Research foundation, had designed a plant microbian fuel cell called plant-e, produces electricity from plants
\cite{plant-e}. The fuel cell work by using the mutual cooperation of bacteria present in the soil and the 
plants (cf. Fig. \ref{pmfc}). 
~\\
} 
}

\section*{Acknowledgement}
AMJ thanks DST, India for financial support (through J. C. Bose National Fellowship).


\end{document}